\documentclass[prl,nobalancelastpage,twocolumn,superscriptaddress]{revtex4-2}

\usepackage[utf8]{inputenc}
\usepackage[american,british]{babel}
\usepackage[T1]{fontenc}
\usepackage[pdftex]{graphicx}  
\usepackage{graphicx, xcolor}
\usepackage{dcolumn}
\usepackage{physics}
\usepackage{braket}
\usepackage{bm}
\usepackage{amsmath,amsthm,amssymb}
\usepackage{color}
\usepackage{verbatim}
\usepackage{ulem}

\usepackage{hyperref}
\hypersetup{
 colorlinks=true,
 linkcolor=blue,
 anchorcolor = blue,
 citecolor = blue,
 filecolor = blue,
 urlcolor = blue
}
\usepackage{dsfont}

\graphicspath{{Images/}}
\newcommand{\rvline}{\hspace*{-\arraycolsep}\vline\hspace*{-\arraycolsep}}

\def \be {\begin{equation}} 
\def \ee {\end{equation}} 
\def \l {\left(} 
\def \r {\right)} 
\def \la {\langle} 
\def \ra {\rangle}

\begin{document}

\title{Energy-filtered quantum states and the emergence of non-local correlations}
\date{\today}

\author{Gianluca Morettini}
\affiliation{Universit\'e Paris-Saclay, CNRS, LPTMS, 91405, Orsay, France.}

\author{Luca Capizzi}
\affiliation{Universit\'e Paris-Saclay, CNRS, LPTMS, 91405, Orsay, France.}

\author{Maurizio Fagotti}
\affiliation{Universit\'e Paris-Saclay, CNRS, LPTMS, 91405, Orsay, France.}

\author{Leonardo Mazza}
\affiliation{Universit\'e Paris-Saclay, CNRS, LPTMS, 91405, Orsay, France.}

\begin{abstract}

Energy-filtered quantum states are promising candidates for efficiently simulating thermal states. We explore a protocol designed to transition a product state into an eigenstate located in the middle of the spectrum; this is achieved by gradually reducing its energy variance, which allows us to comprehensively understand the crossover phenomenon and the subsequent convergence towards thermal behavior. We introduce and discuss three energy-filtering regimes (short, medium and long), and we interpret them as stages of thermalization. We show that the properties of the filtered states are locally indistinguishable from those of time-averaged density matrices, routinely employed in the theory of thermalization. On the other hand, non-local quantum correlations are generated in the medium regimes and are witnessed by the R\'enyi entanglement entropies of subsystems, which we compute via replica methods. Specifically, two-point correlation functions break cluster decomposition and the entanglement entropy of large regions scales as the logarithm of the volume during the medium filter time.

\end{abstract}

\maketitle

\paragraph{Introduction ---}

In a seminal article, M.C.~Ba\~{n}uls, D.~Huse and J.I.~Cirac investigated
how much entanglement is necessary to reduce the energy fluctuations of a quantum state in the middle of the spectrum of a many-body system~\cite{bhc-20}. 
The question is natural: 
on the one hand, product states have no entanglement but extensive energy fluctuations, on the other hand, the exact eigenstates of Hamiltonians display extensive entanglement entropy. 
To understand the crossover, 
they introduced the protocol of \textit{energy filtering} (analogous to Refs.~\cite{Hastings-04,Haegeman-13}), 
which progressively reduces the energy fluctuations of an initial product state.
Remarkable was the discovery of intermediate regimes where energy variance shrinks to zero while the entropies grow logarithmically in the system size~\cite{bhc-20,rca-23}, opening the way to reproducing efficiently thermal properties via pure 
states~\cite{sbc-21,ccb-21,ltbc-23,ibc-23,lgc-24,sbck-23,Hemery-23,gsd-23,skngg-14,Jiaozi-22,pfk-24}.

A protocol analogous to filtering is unitary dynamics as, in both cases, purity is preserved but, at late times, the state becomes locally indistinguishable from a thermal one. Quantum correlations spreading across the system during real-time evolution have been thoroughly characterized in order to gain a deeper understanding of the thermalization process. On the other hand, fundamental questions regarding quantum correlations in energy-filtered quantum states (EFQS) are still unanswered. In addition to theoretical interest, clarifying these issues is crucial for understanding the extent to which EFQS can accurately capture thermal properties and to discern features that are instead related to non-thermal spurious effects.

In this Letter, we give a detailed characterization of the behavior of both local observables, their correlations and entanglement measures during filtering. Remarkably, we find that the EFQS exhibits non-local quantum correlations of local observables at arbitrarily distant points. We further develop a technique, based on replica methods, to compute the entanglement entropies of large regions, which complements the information on correlation functions. We find that the intermediate regime of the protocol, characterized by small energy variance and entanglement, is unavoidably accompanied by the violation of the cluster decomposition principle, which is an unorthodox non-thermal feature.
Our approach applies to generic $d$-dimensional systems and provides model-independent predictions 
that we numerically validate 
on a non-integrable spin chain. 

\paragraph{Energy filters ---}

We begin by recalling a few known results on EFQS.
We consider a local Hamiltonian $H $ and a product 
state $\ket{\Psi_0}$ that is not an energy eigenstate and such that $E_{\Psi_0}=\bra{\Psi_0} H \ket{\Psi_0}$ lies in the middle of the energy spectrum.
We construct the EFQS as follows:
\begin{equation}\label{eq:filtered_state}
\ket{\Psi_\tau} = 
\frac{1}{\sqrt{\mathcal Z(\tau)}} 
\exp\left( -\frac{(H-E_{\Psi_0})^2 }{4} \tau^2
\right)
\ket{\Psi_0};
\end{equation}
the operator acting on $\ket{\Psi_0}$ is the \textit{energy filter} and $\tau$ the \textit{filter time};
$\mathcal Z(\tau)$ ensures the normalisation of the filtered state as $\braket{\Psi_\tau|\Psi_\tau}=1$.
The energy variance of $\ket{\Psi_\tau }$ decreases as $\tau$ increases, and $\ket{\Psi_\tau}$ interpolates between the initial product state $\ket{\Psi_0}$ and a state with reduced energy fluctuations maintaining the same energy. 
Without loss of generality, 
we can assume $E_{\Psi_0}=0$ (the Hamiltonian can be shifted by a constant).

The energy distribution of $\ket{\Psi_\tau}$ can be determined under weak assumptions.
For a product state $\ket{\Psi_0}$,
all the cumulants of $H$ are extensive, the central limit theorem holds, and the energy distribution of $\ket{\Psi_0}$ is Gaussian in the large volume $V$ (that is, the number of sites) limit~\cite{hmgh-04}. This means that for a typical eigenstate $\ket{E}$ with energy $E$ the scalar product
$
|\la E | \Psi_0 \ra|^2  \propto \exp \l -\frac{E^2}{2\varepsilon_2 V}\r,
$
where $\varepsilon_2 V\equiv \Delta H^2_{0}$ is the energy variance of $\ket{\Psi_0}$. From Eq.~\eqref{eq:filtered_state} we can compute the energy distribution of the filtered state
$ |\la E | \Psi_\tau \ra|^2 \propto \exp\l -\frac{E^2\tau^2}{2}\r |\la E | \Psi_0 \ra|^2 $,
so that the variance of $\ket{\Psi_\tau}$
is~\cite{bhc-20}
\begin{equation}
\label{eq:En_variance}
\Delta H^2_\tau
\simeq \frac{1}{ \tau^2 + \frac{1}{\varepsilon_2 V}}.
\end{equation}

\paragraph{Filtering regimes --- }

The formula in Eq.~\eqref{eq:En_variance} suggests the identification of three filtering regimes. We list them here and anticipate some of the results derived below.
At \textit{short filter time}, $\tau \sim \frac{1}{\sqrt{V}}$, the energy variance is extensive and the expectation value of local observables is close to the initial value; entanglement starts to build up, but the state remains a standard area-law state.
At \textit{medium filter time}, $\tau \sim O(1)$, the energy variance does not scale anymore with the size of the system. 
Bipartite entanglement entropies become significant and have a universal scaling as $\frac{1}{2} \log V_A$, where $V_A$ is the volume of the smallest region $A$
(reminiscent of the logarithmic behavior found in long-range systems in Refs.~\cite{lp-20,Pappalardi-18}).
Here, a new phenomenology appears: the state breaks the clustering condition and quantum correlations become highly non-local.
Finally, at \textit{long filter times}, when $\tau$ increases with $V$,
local observables attain values that are independent of $\tau$. The bipartite entanglement entropy scales as the volume $\sim V_A$ and two-point correlation functions satisfy again a clustering condition; the state can be considered as thermal. 

\paragraph{Local observables ---}\label{sec:Loc_Obs}

The study of EFQS is not straightforward because $\ket{\Psi_\tau}$ in Eq.~\eqref{eq:filtered_state} is issued from a non-local and non-unitary evolution~\footnote{For instance, $H^2$ couples arbitrarily distant points, and the operator $\exp\l -\frac{H^2 \tau^2}{4}\r$ in~\eqref{eq:FT_trick} is non-unitary.}.
Our approach is based on the existence of a deep link with the process of thermalization that we detail below.

We begin with a representation of $\ket{\Psi_\tau}$ obtained by Fourier-transforming the energy filter (see also Refs.~\cite{sstfm-17,Osborne-06,bhc-20})
\begin{equation}\label{eq:FT_trick}
\ket{\Psi_\tau} \propto \int^{\infty}_{-\infty} 
\hspace{-0.15cm}
dt' \lambda_\tau(t')e^{-iHt'}\ket{\Psi_0}, \; \;
\lambda_\tau(t) =\exp \bigl( -\tfrac{t^2}{\tau^2} \bigr).
\end{equation}
Eq.~\eqref{eq:FT_trick} provides a link between the filtered state and the time-evolving one $\ket{\Psi_0(t)} =e^{- i H t} \ket{\Psi_0}$.
We now consider a local observable $\mathcal{O}$~\footnote{We refer to operators acting on a finite number of sites, and limits of their sequences in the operator norm. The interested reader can find a precise definition in Refs. \cite{Robinson-67,br-87}.} in the Heisenberg picture, $\mathcal{O}(t) = e^{iHt}\mathcal{O}e^{-iHt}$. 
Formally, its expectation value in $\ket{\Psi_\tau}$, which we denote by $\langle \ldots \rangle_\tau$, reads
\begin{equation}\label{eq:O_time}
\la \mathcal{O}(t)\ra_\tau = \frac{\int d\tilde{t}_1 dt_1 \lambda_\tau^*(\tilde{t}_1)\lambda_\tau(t_1) \la e^{iH(\tilde{t}_1-t_1)}\mathcal{O}(t-t_1)\ra_0}{\int d\tilde{t}_1 dt_1 \lambda_\tau^*(\tilde{t}_1)\lambda_\tau(t_1) \la e^{iH(\tilde{t}_1-t_1)}\ra_0}.
\end{equation}
\begin{figure}[t]
 \includegraphics[width=\columnwidth]{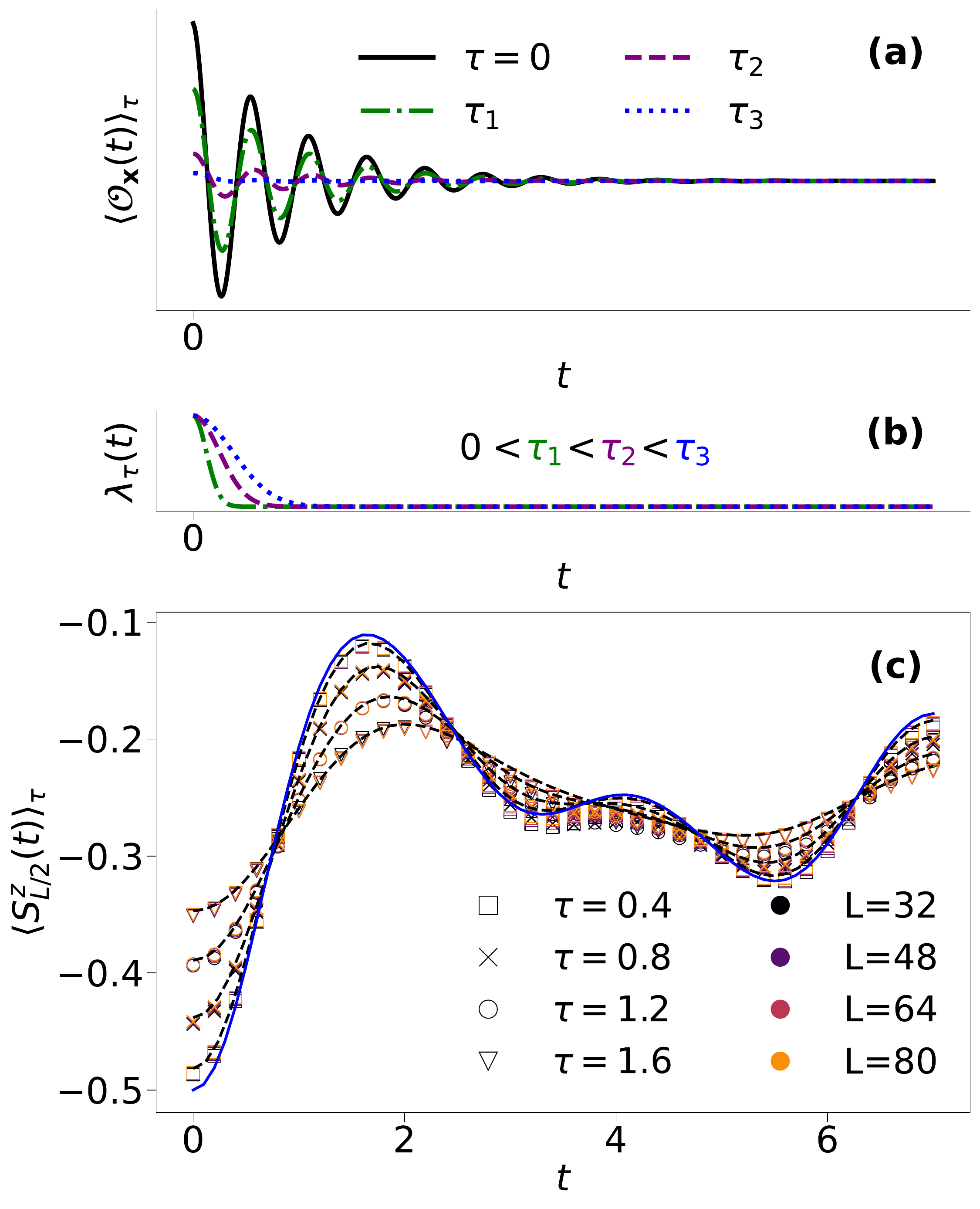}
 \caption{
 A sketch of $\langle \mathcal O_\mathbf{x}(t) \rangle_\tau$ as a function of $t$ for different values of $\tau$ is shown in panel (a). 
 The data are produced using Eq.~\eqref{eq:O_med_t} 
 and we plot the profile of the filter kernel $\lambda_\tau(t)$ that we employed in panel (b). 
 In panel (c) we show actual numerical results for the model discussed at the end of the letter in Eq.~\eqref{eq:H_ising}. 
 The markers represent the numerical data obtained for the numerically-simulated filtered state, while the 
 blue line shows the expectation value of the observable
 for the unfiltered state $\ket{ \Psi_0(t)}$.
 The comparison with the black dotted lines, which are evaluated via the right-hand side of Eq.~\eqref{eq:O_med_t}, is excellent at several system sizes $L$.
 }
 \label{Fig:Sketch}
\end{figure}
%
We point out that,
in the limit of large system $V\rightarrow \infty$ with $t,\tau$ fixed, this can be simplified: 
On the one hand, at the denominator
the expectation value $\langle e^{ iH(\tilde{t}_1-t_1)} \rangle_0$ is the Loschmidt echo, or return amplitude, which scales as $e^{ VF(t_1-\tilde{t}_1)}$ \footnote{This asymptotic behavior comes from the extensivity of the cumulants of $H$ in the large volume limit.}, with $F(t)$ a function that is analytic in a neighborhood of $t=0$, when $\ket{\Psi_0}$ is a product state \cite{mpk-13,Pozsgay-13}.
 On the other hand, such an exponential localisation in time characterises also the numerator $\langle e^{iH(\tilde{t}_1-t_1)}\mathcal{O}(t-t_1)\rangle_0$, which turns out to scale as $e^{ VF(t_1-\tilde{t}_1)} \la \mathcal{O}(t-t_1)\ra_0$.
We can then perform the integrals over $\tilde{t}_1$ in Eq.~\eqref{eq:O_time}, which are marked by a saddle point contribution localized at $\tilde{t}_1\simeq t_1$. This gives the first main result:
\begin{equation}\label{eq:O_med_t}
\langle \mathcal{O}(t)\rangle_\tau \simeq  \frac{\int dt_1 |\lambda_\tau(t_1)|^2\langle\mathcal{O}(t+t_1)\rangle_0}{\int dt_1 |\lambda_\tau(t_1)|^2}.
\end{equation}
Hence, the EFQS $\ket{\Psi_\tau}$ is locally indistinguishable from the time-averaged mixed state~\footnote{
While we focused on Gaussian filters, other equivalent choices are considered in the literature, as for example the box function $\lambda'_\tau(t) = \chi_{[-\tau/2,\tau/2]}(t)$. We do not expect significant qualitative differences with other choices of the filter function.}~\cite{Fagotti-19,ccb-21}
\begin{equation}
\label{eq:time-averaged:incoherent}
\rho(\tau) \propto \int dt_1 \, |\lambda_\tau(t_1)|^2 \, 
\ket{\Psi_0(t_1)} \hspace{-0.11cm} \bra{\Psi_0(t_1)}.
\end{equation}
Eqs.~\eqref{eq:O_med_t} and~\eqref{eq:time-averaged:incoherent} are a powerful tool for linking the physics of EFQS to the theory of thermalization, as we discuss below---see also Fig.~\ref{Fig:Sketch}.
Let us consider for simplicity $t=0$ in Eq.~\eqref{eq:O_med_t} and a local observable $\mathcal{O}$.
For small $\tau$, $\langle \mathcal O(0) \rangle_\tau$ is dominated by the initial transient dynamics of $\langle \mathcal O(t_1) \rangle_0$. 
For large $\tau$, it is dominated by times $t_1$ larger than the observable's relaxation time, and, thus,
$\langle \mathcal O(0) \rangle_\tau$ converges to its thermal value.

In the rest paper, we will characterize the correlations developed during the filtering giving predictions for the two-point functions and the entanglement entropy of large regions.

\paragraph{Two-point correlations ---}
Let us focus on short or medium filter times $\tau$ and consider two operators $\mathcal{O}_{\mathbf{x}}$ and  $\mathcal{O}_{\mathbf{y}}$, localized at $\mathbf{x},\mathbf{y}$, at a distance large enough to cluster in the state $e^{-iH
t}\ket{\Psi_0}$ with $|t|\lesssim \tau$. Eq.~\eqref{eq:O_med_t} can then be simplified as follows
\begin{equation}\label{eq:LRO}
\langle \mathcal{O}_{\mathbf{x}}\mathcal{O}_{\mathbf{y}}\rangle_{\tau,c} \simeq \frac{\int dt_1 |\lambda_\tau(t_1)|^2 \la \mathcal{O}_{\mathbf{x}}(t_1)\ra^2_0}{\int dt_1 |\lambda_\tau(t_1)|^2} - 
\langle \mathcal O_{\bf x}\rangle_\tau ^2,
\end{equation} 
where translational invariance, that implies $\langle \mathcal O_{\bf x}\rangle_\tau = \langle \mathcal O_{\bf y}\rangle_\tau$, has been employed. 
This correlator satisfies the clustering condition if, for large enough $|\mathbf x - \mathbf y|$, it is equal to zero. Concerning the integral, since the initial state satisfies the clustering condition, it holds $\la \mathcal{O}_{\bf x}(t) \mathcal{O}_{\bf y}(t) \ra_0 \simeq \la \mathcal{O}_{\bf x}(t)\ra_0 \la \mathcal{O}_{\bf y}(t) \ra_0$ for sufficiently distant $\bf{x}$ and $\bf{y}$.

\begin{figure}[t]
 \includegraphics[width=\columnwidth]{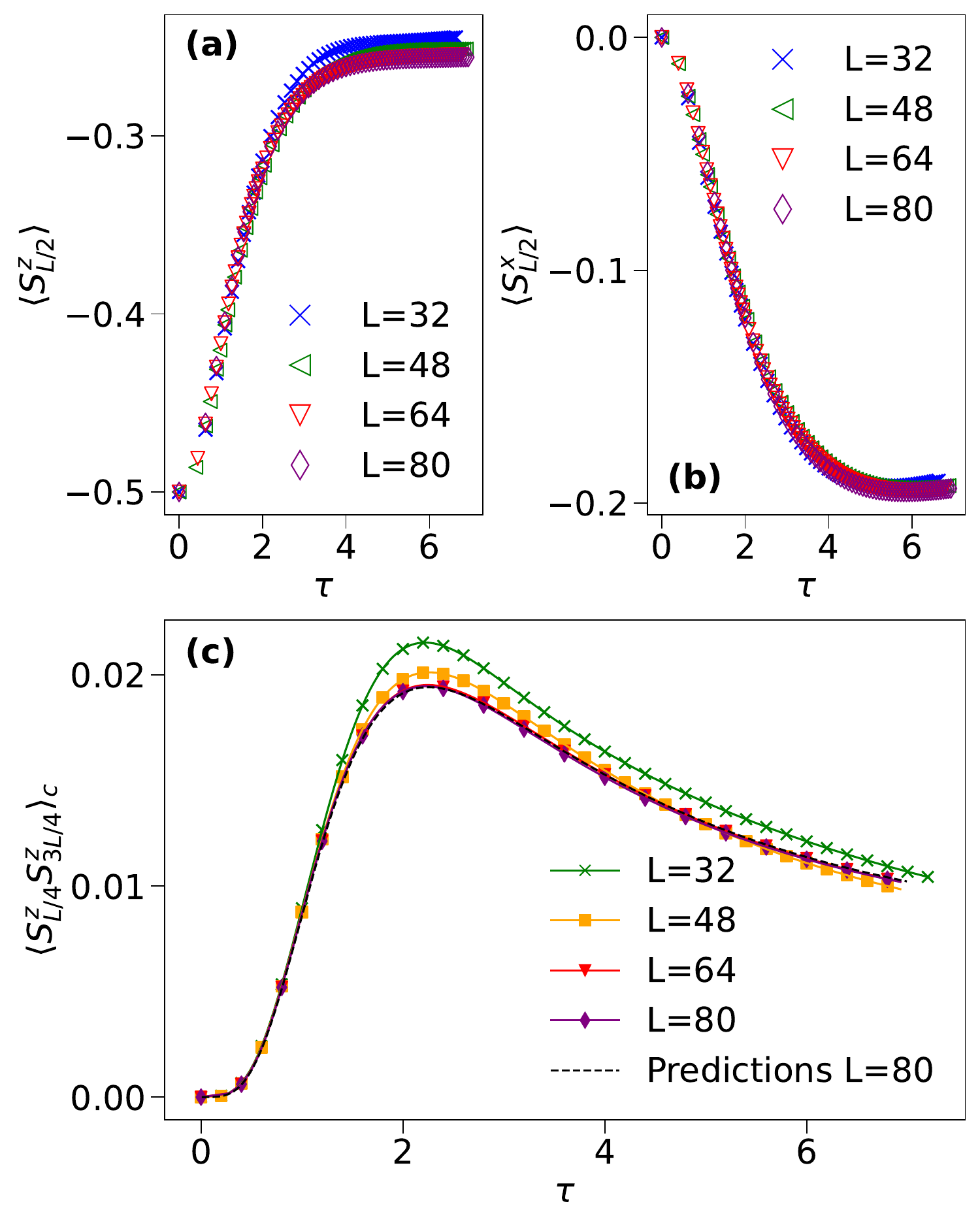}
\caption{Panels (a) and (b) show the evolution of $\langle S^z_{j=\frac{L}{2}} \rangle_\tau$ and $\langle S^x_{j=\frac{L}{2}} \rangle_\tau$, as function of $\tau$. In (c) we show results for the connected correlator $\langle S^z_{j=\frac{L}{4}} S^z_{j'=\frac{3L}{4}}\rangle_{\tau,c}$, as function of $\tau$. We compare numerical data for $L=32, 48, 64, 80$ with the right-hand side of Eq.~\eqref{eq:LRO} evaluated for $L=80$.}\label{Fig:Time:Tau:Link}\label{Fig:Breaking:Clustering}
\end{figure}

At short filter times, where $|\lambda_\tau(t)|^2$ localizes at $t\simeq~0$, $\langle \mathcal{O}_{\mathbf{x}}\mathcal{O}_{\mathbf{y}}\rangle_{\tau,c}$  vanishes because the initial state $\ket{\Psi_0}$ is a product state. For medium $\tau$, instead, the right-hand side of Eq.~\eqref{eq:LRO} is nonzero, provided $\la \mathcal{O}_\mathbf{x}(t)\ra_0$ depends on $t$ explicitly--- we will provide numerical evidence in Fig.~\ref{Fig:Breaking:Clustering}. 
In that regime, therefore, the state $\ket{\Psi_\tau}$ breaks the clustering condition and its correlations become non-local, as it happens for the density matrix $\rho(\tau)$ in Eq.~\eqref{eq:time-averaged:incoherent}.
This is consistent with the fact that the energy filter introduced in Eq.~\eqref{eq:filtered_state} comes from a non-local time evolution: for instance, a local unitary evolution does not develop non-local correlations, as a consequence of the Lieb-Robinson bound.
Finally, for larger $\tau$, we can show that $\lim_{\tau \to \infty} \lim_{|\mathbf x - \mathbf y| \to \infty} \lim_{L \to \infty} \langle \mathcal O_{\bf x} \mathcal O_{\bf y} \rangle_{\tau, c} =0$ under the assumption that $\langle \mathcal O_{\bf x}(t)\rangle_0$ converges identically to its stationary value $\overline{\langle \mathcal O_{\bf x} \rangle_0}$, where the bar denotes time average, in the limit $t \to \infty$.
Indeed, recalling that in the $\tau \to \infty$ limit filtering is equivalent to averaging over the entire time dynamics, both terms appearing in the definition of Eq.~\eqref{eq:LRO} are equal to $\overline{\langle \mathcal O_{\bf x} \rangle_0}^2$, and thus subtract identically.



\paragraph{Entanglement of bipartitions --- } \label{sec:entanglement}

We now investigate quantum correlations between spatial regions through the lens of entanglement. We recall that, given a quantum state and denoting by $\rho_A$ its reduced density matrix with respect to a region $A$ with volume $V_A$, its 
$n$-th R\'enyi entropy reads
$
S_n(A) \equiv (1-n)^{-1}\log \text{Tr}\l \rho^n_A\r
$
and the von Neumann entropy is $S_1(A) \equiv - \text{Tr}\l \rho_A \log \rho_A\r$.
We compute the R\'enyi entropies $S_{n,\tau}(A)$ ($n\geq 2$) for the filtered state $\ket{\Psi_{\tau}}$ with the replica method, which requires to compute $\text{Tr}\l \rho^n_A\r$ as a certain partition function between $n$ replicas; we then use the replica trick to compute the von Neumann entropy, i.e., we perform the analytical continuation over $n \in \mathbb R_{>0}$ and take the limit $n\rightarrow 1^+$~\cite{cc-04}.
After introducing the twist operator $\mathcal{T}_A$, that acts as a replica permutation $j\rightarrow j+1$ (with $j=1,\dots,n$ a replica index) within the region $A$,
we can represent the moments of $\rho_A$ as
$
\text{Tr}\l \rho^n_A\r = \, ^{n}\hspace{-0.1cm}\bra{\Psi}\mathcal{T}_A\ket{\Psi}^{n}$,
with $\ket{\Psi}^{n}\equiv \ket{\Psi}^{\otimes n}$ the state of the system replicated $n$ times~\cite{ccd-07,cc-09,cd-11,hms-14}. 
Details on the actual calculations are reported in the Supplementary Material (SM); here we focus on the results.

At short filter times $\tau \sim 1/\sqrt{V}$, entanglement starts to build up quickly, and we find
\begin{equation}
S_{n,\tau}(A)  - S_{n,0}(A) = f_{\mathrm {sft},n}\left( \sqrt{\varepsilon_2 V} \tau, \frac{V_A}{V}\right).
\end{equation}
In the case of a product state $S_{n,0}(A)$ is zero, but the result holds more generally for an initial area-law state. The explicit form of $f_{\mathrm{ sft},n}$ is universal and it does not depend on the details of the Hamiltonian; it is reported in Eqs.~\eqref{eq:Sn_sh_filter} and \eqref{eq:det_anal} of the SM. 
In particular, it first grows quadratically as $f_{\mathrm {sft},n} \sim V \tau^2$, whereas asymptotically the behaviour is logarithmic, $f_{\mathrm {sft},n} \sim \log ( \sqrt{V} \tau)$. After this transient, at a filter time scaling as $\tau \sim 1/\sqrt V$, the entropy reaches a value proportional to the logarithm of $V_A$.
This result is non-trivial as a priori one would expect that a non-local evolution saturates the quantum correlations and produces an entanglement of bipartition scaling as $V_A$.

The medium filter time regime takes place after this saturation and we obtain 
\begin{equation}\label{eq:Renyi_nbig}
S_n(A)-S_{n,0}(A) \simeq \frac{1}{2}\log V_A + g_{\mathrm{mft},n}(\tau) +  \ldots .
\end{equation}
The function $g_{\mathrm{mft},n}(\tau)$, represents the most important contribution that depends on $\tau$. In particular, the behaviour above is found for $\tau$ shorter than
the time that is necessary for the subsystem to thermalize. 

The explicit expression of $g_{\mathrm{mft},n}(\tau)$ depends on the model, but its asymptotic behavior in $\tau$ has general features; we find a strong dependence on the order of the R\'enyi entropy, a situation which is rather uncommon:
\begin{equation}
\label{eq:gmft:cases}
 g_{\mathrm{mft},n}(\tau) \simeq 
 \begin{cases}
  \frac{n}{n-1}\log \tau, & n > 1;\\
  \frac{1}{\sqrt{2 \pi}} \Gamma_1 \, |\partial A|\tau, & n = 1;\\
  \frac{1-n }{8n} \Gamma^2_n \, |\partial A|^2 \tau^2, & 0< n < 1.\\
 \end{cases}
\end{equation}
Here, the growth of entropy under unitary dynamics is assumed to be linear, and $\Gamma_n$ is related to its rate via $S_n(A) \simeq \Gamma_n \, |\partial A| t$. 
The linear growth of $S_1(A)$ as a function of $\tau$ in Eq. \eqref{eq:gmft:cases} has to be compared with the slower logarithmic growth observed for $n>1$, and it is compatible with the rigorous upper bound found in Ref.~\cite{bhc-20} with different methods for one-dimensional systems. We mention that a similar drastic change of behaviour of the R\'enyi entropy close to $n=1$ has been found also in a different context in Refs.~\cite{fzb-23,rpv-19}. 

The R\'enyi entropies for $0<n<1$ are fundamental in the context of tensor networks; in particular, it is known that if $S_n(A)$ is bounded by $c \log V_A$, then an efficient MPS representation exists~\cite{cpsv-21}. Our result in Eq.~\eqref{eq:gmft:cases}, therefore,  suggests that the filtered state $\ket{\Psi_\tau}$ can be efficiently simulated with a tensor-network algorithm up to a filter time of order
$
\tau \sim \sqrt{\log V}.
$
This would allow to simulate classically a state with logarithmic entanglement entropy and
an energy variance decaying as $\Delta H^2_\tau \sim \frac{1}{\log V}$. A rigorous proof of a similar statement has been provided in Ref.~\cite{rca-23} with different methods for one-dimensional systems. 

We point out that Ref.~\cite{gmm-23} found asymptotic eigenstates with subextensive entanglement entropy and energy variance $\Delta H^2  \sim \frac{1}{L^2}$: such a scenario cannot be recovered via the filter protocol for generic systems because $\tau \sim V$ is a time scale compatible with the thermalization time, when every region has thermalized and an extensive entropy is observed.

\paragraph{Mutual information ---}

We remark that the prefactor ($1/2$) of the logarithmic growth of the entropy as a function of the volume in Eq.~\eqref{eq:Renyi_nbig} holds for both connected and disconnected regions. In particular, the mutual information of two distant regions $A,B$ of size $V_A,V_B=V_A$ can be estimated as
\be\label{eq:MInfo}
I(A,B) \equiv S_1(A)+S_1(B)-S_1(A\cup B) \simeq \frac{1}{2}\log V_A +\dots
\ee
up to $O(1)$ terms (finite for $\tau\neq0 $ fixed and $V_A$ large). Therefore, the mutual information does not decay to zero with distance, as it happens in the ground states of critical systems~\cite{fps-09}. Similar properties are also found in some exact scars (see Ref.~\cite{dppgp-22}), and they are ultimately related to the breaking of the clustering condition.
This is the entanglement version of the non-locality of the state at medium filter time that has been pointed out when looking at two-point correlation functions. Finally, for large filter times, where the thermal state is expected to be approached and the long-range correlations are absent, the mutual information between distant regions should eventually vanish.

\paragraph{Numerical simulations --- }

We benchmark our predictions against numerical results in a one-dimensional quantum spin chain of length $L$. We consider the spin-1/2 Ising model with both longitudinal and transverse magnetic field
\be \label{eq:H_ising}
H=-J\sum_{j=1}^{L-1}{S^x_jS^x_{j+1}}+h_x\sum_{j=1}^{L}{S^x_j}+h_z\sum_{j=1}^{L}{S^z_j};
\ee
with open boundary conditions, and we choose $J=1, h_x=1.2, h_z=0.8$.  
We take
$
\ket{\Psi_0} = \ket{\uparrow\downarrow\uparrow\downarrow\ldots\uparrow\downarrow}
$, which lies in the middle of the spectrum with $E_{\Psi_0}=0$ and has extensive energy variance $\Delta H^2_0 = (J^2/16 + h_x^2/4) L$.
In order to numerically implement the energy filter $e^{ - \frac{H^2\tau^2}4}$ we use
 a matrix-product state representation of the quantum state combined with the Time-Dependent Variational Principle (TDVP)~\cite{review_TDPV, Cirac_1TDVP, Haegeman_2TDVP}, employing the ITensors library \cite{ITensor, ITensor-r0.3}. 
We follow Ref.~\cite{Goto_TDVP} and we first apply the 2-TDVP  until a chosen bond-link $\chi=450$ is reached; subsequently, we employ the 1-TDVP algorithm at fixed bond link. This procedure represents a compromise between the computational efficiency offered by 1-TDVP, and the mitigation of projection errors inherent to 2-TDVP.
This allows us to reliably simulate chains of up to $L=80$ sites up to filter time $\tau = 7$. 

We first assess the validity of Eq.~\eqref{eq:O_med_t}. In Fig.~\ref{Fig:Sketch}(c) we plot the numerical data obtained for $\langle \mathcal O (t) \rangle_\tau$, with $\mathcal O = S^z_{j = \frac L2}$. Black dashed curves are obtained by processing, according to Eq.~\eqref{eq:O_med_t}, the unitary time-evolution result $\langle \mathcal O (t) \rangle_0$, which is plotted as a solid blue line. 
The comparison with the direct simulation of $\langle \mathcal O(t) \rangle_\tau$ using the filter algorithm is excellent.

We subsequently investigate the behaviour of local observables and of two-point connected correlation functions of distant points, studying $\langle S^z_{j= \frac L2}\rangle$, $\langle S^x_{j= \frac L2} \rangle$ and $\langle S^z_{j=\frac L4}  S^z_{j'=\frac {3L}4} \rangle_{\tau,c}$
and plot the results in Fig.~\ref{Fig:Breaking:Clustering}.
Local observables have a significant dependence on the filter time up to $\tau \sim 4$, after which they show significant saturation effects: this plateau is non thermal, since the Gibbs ensemble associated with $\ket{\Psi_0}$ is the infinite-temperature state~\footnote{We remark that the temperature of a state is determined by the expectation value energy density; in particular, the latter is equal to that of the corresponding thermal state.} and the corresponding expectation value of the Pauli matrices is zero. We believe that its origin comes from the average of the fast transient oscillations present in real-time dynamics and shown in Fig.~\ref{Fig:Sketch}; we expect that the thermal value, that is $0$, is eventually approached at larger times $t$ in the unitary evolution, and therefore at larger $\tau$ in the filtering protocol (see SM for details).

As anticipated, the connected correlator takes the value $0$ at $\tau=0$, increases towards a maximal value, and eventually decreases for longer values of $\tau$. The curves obtained at various $L$ are compatible with a collapse as $L$ increases. We also evaluate numerically for $L=80$ the right-hand side of Eq.~\eqref{eq:LRO}, obtained from the unitary dynamics of the model,  finding good agreements with the previous curves. Note that while the connected correlation function is expected to converge eventually to zero, as $\tau\rightarrow \infty$, the decreasing trend is slow.

\begin{figure}[t]
\includegraphics[width=\columnwidth]{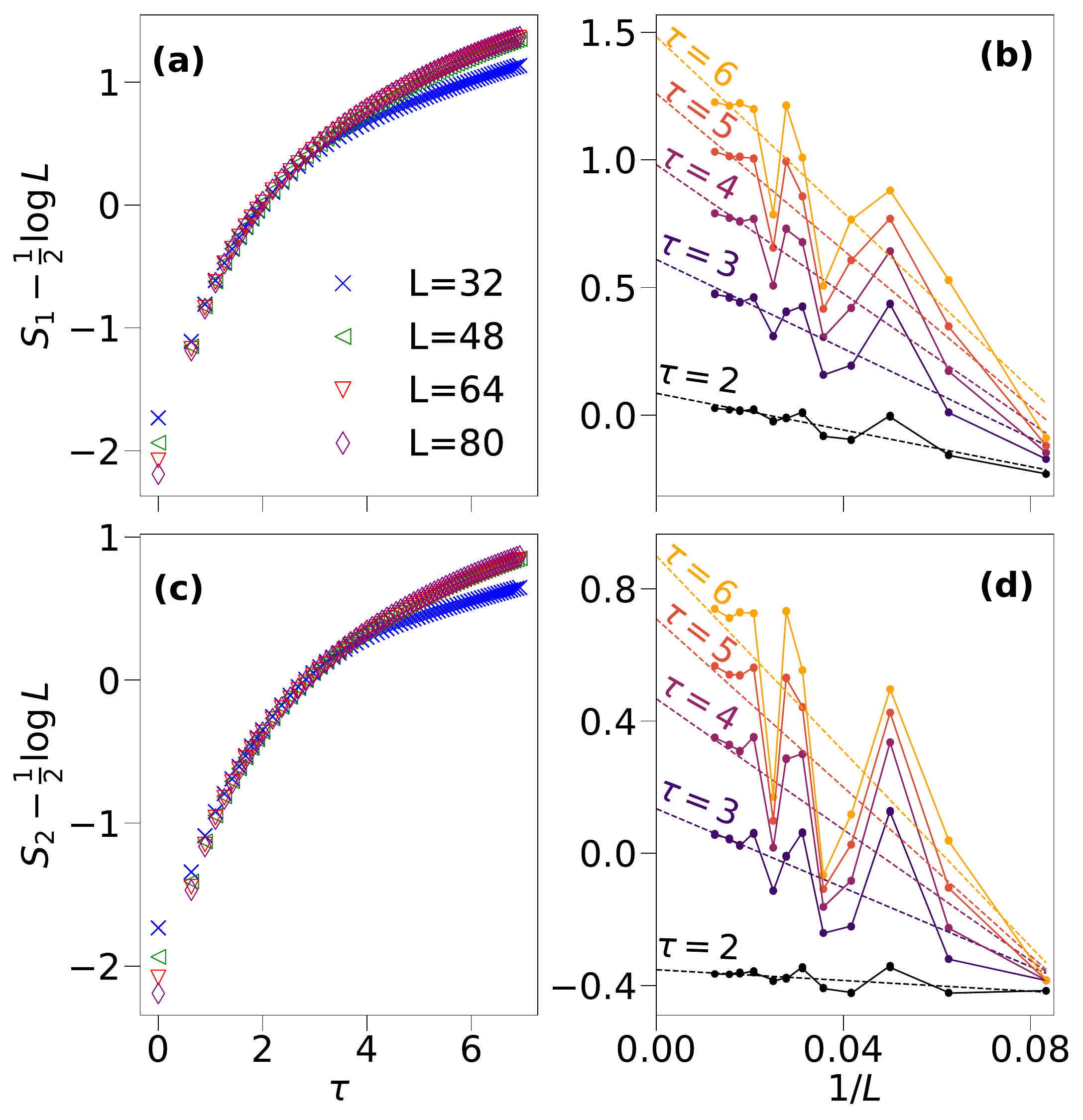}
 \caption{$S_n-1/2 \log L$ ($n=1,2$) for the half-chain bipartition as a function of $\tau,L$. The data against $\tau$ (panels (a) and (c)) show compatibility with a collapse as $L$ increases. In (b) and (d) we show the results as a function of $1/L$ for different values of $\tau$ and extrapolate them in the limit $L \to \infty$. The extrapolated value is finite, which supports the anticipated large-$L$ behaviour $S_{1,2} \sim \frac 12 \log L$.}\label{Fig:S1:S2:MFT}
\end{figure}

Finally,
we study the R\'enyi entropies ($n=1,2$) of half-chain in the medium filter time as a function of $\tau$ and system size $L$;  the results are plotted in Fig.~\ref{Fig:S1:S2:MFT}. Our data show compatibility with a collapse of $S_n - \frac{1}{2}\log L$ against $\tau$, as predicted by Eq.~\eqref{eq:Renyi_nbig}. Finite-size deviations are displayed in panels (b) and (d), and a slowly-decaying oscillating behaviour as a function of $L$ is found. For such small values of $\tau$, it is impossible to attempt a study of the asymptotic behaviours discussed above.

\paragraph{Conclusions ---}

We have shown that the filtering protocol generates non-local correlations in an intermediate filter-time regime. 
We provide analytical predictions for entanglement entropies and its filtering dynamics that address fundamental questions regarding the simulability of thermal properties via quantum pure states, and that have been thoroughly validated through numerical simulation. 

An interesting direction regards the possible role of conserved charges, or integrability, in the filtering. 
Non-thermal states, say generalized Gibbs ensembles, are then expected to arise eventually after filtering. We defer the examination of these generalizations to future investigations.

\paragraph{Acknowledgements ---}
We thank M.~C.~Ba\~{n}uls, A.~De~Luca, R.~Fazio and J.~Wang for enlightening discussions.
LM~acknowledges discussion with L.~Gotta and S.~Moudgalya on a previous related work.
LC and MF acknowledge support from ERC Starting grant 805252 LoCoMacro. 
This work has benefited from a State grant as part of France 2030 (QuanTEdu-France), bearing the reference ANR-22-CMAS-0001 (GM),
and is part of HQI (www.hqi.fr) initiative,  supported by France 2030 under the French National Research Agency award number ANR-22-PNCQ-0002 (LM).

\bibliography{bibliography}

\onecolumngrid
\appendix

\newpage 

\setcounter{secnumdepth}{2}

\begin{center}
\begin{large}
\textbf{Supplementary Material for\\
``Energy-filtered quantum states and the emergence of non-local correlations''}
\end{large}
\end{center}

\section{R\'enyi entropies and replica trick}\label{app:Renyi_replica}

In this section, we characterize the R\'enyi entropy of the filtered state $\ket{\Psi_\tau}$ via replica trick. Given a subsystem $A$, we first express the $n$th moment of its RDM as 
\be
\begin{split}\label{eq:Exp_twist}
& \text{Tr}\l \rho^n_A\r = ^{n}\bra{\Psi_\tau}\mathcal{T}_A\ket{\Psi_\tau}^{n} = \frac{1}{\l \int d\tilde{t}_1 dt_1 \lambda(\tilde{t}_1)\lambda_\tau(t_1)  \la e^{iH(\tilde{t}_1-t_1)}\ra_0\r^n}\times\\
&\int d\tilde{t}_1 dt_1\dots d\tilde{t}_n dt_n \lambda_\tau(\tilde{t}_1)\dots \lambda_\tau(\tilde{t}_n)\lambda_\tau(t_1)\dots \lambda_\tau(t_n) \times ^{n}\bra{\Psi_0} e^{i\sum_j H_j \tilde{t}_j} \mathcal{T}_A e^{-i\sum_j H_j t_j}\ket{\Psi_0}^n,
\end{split}
\ee
with $H_j$ the Hamiltonian of the $j$-th replica. The expression above, exact at finite size, boils down to a computation of an integral in $2n$ variables. Simplifications occur in the limit of large regions, and one can exploit the exponential localization of the integrand (analogous to the Loschmidt echo case, discussed in the main text). For instance, $^{n}\bra{\Psi_0} e^{i\sum_j H_j \tilde{t}_j} \mathcal{T}_A e^{-i\sum_j H_j t_j}\ket{\Psi_0}^n$ gets a contribution from the region $A$ associated with the replica shift: this gives localization around $t_{j-1}\simeq \tilde{t}_j$. Another contribution comes from the complementary region, and it gives localization around $t_{j}\simeq \tilde{t}_j$. We perform a quadratic approximation around the localization points, relevant to evaluate the integral in Eq.~\eqref{eq:Exp_twist}, and we get 
\be\label{eq:Exp_twist1}
\begin{split}
&^{n}\bra{\Psi_0} e^{i\sum_j H_j \tilde{t}_j} \mathcal{T}_A e^{-i\sum_j H_j t_j}\ket{\Psi_0}^n  \simeq \exp \l -\frac{\varepsilon_2 (V-V_A)}{2} \sum_j (t_j-\tilde{t}_j)^2-\frac{V_A\varepsilon_2}{2}\sum_j (t_{j-1}-\tilde{t}_{j})^2\r \times {}^n\bra{\Psi_0}\mathcal{T}_A(t_1) \ket{\Psi_0}^n.
\end{split}
\ee
Here, ${}^n\bra{\Psi_0}\mathcal{T}_A(t_1) \ket{\Psi_0}^n$ is the $n$th moment of the RDM in the state $\ket{\Psi_0(t_1)}$, and it is the only quantity which contains detail on the underlying model.

We first discuss the short filter time $\tau \sim \frac{1}{\sqrt{V}}$ and for convenience we define the rescaled variable
\be\label{eq:t_tilde}
\tilde{\tau} = \sqrt{V}\tau.
\ee
The integral in Eq.~\eqref{eq:Exp_twist} is a Gaussian integral over $2n$ variables. 
Because of  the fast decay of $\lambda_\tau(t)$ as a function of $t$, the integrand is localized around $t_j\simeq\tilde{t}_j\simeq0$; in this regime we can therefore safely replace $^n\bra{\Psi_0}\mathcal{T}_A(t_1) \ket{\Psi_0}^n$ by $^n\bra{\Psi_0}\mathcal{T}_A(0) \ket{\Psi_0}^n$ inside \eqref{eq:Exp_twist}. The latter is $1$ for a product state, but the forthcoming discussion will hold true for area-law states as well. We introduce the $2n$-dimensional vector
\be
\mathbf{t} = \begin{pmatrix}t_1\\ \dots\\ t_n\\ \tilde{t}_1\\ \dots\\ \tilde{t}_n\end{pmatrix},
\ee
and we express the numerator of \eqref{eq:Exp_twist} as
\be
{}^n\bra{\Psi_0}\mathcal{T}_A(0) \ket{\Psi_0}^n \times \int d\mathbf{t} \exp\l -\frac{1}{2}V \mathbf{t}^T \mathbf{M}_n \mathbf{t} \r = {}^n\bra{\Psi_0}\mathcal{T}_A(0) \ket{\Psi_0}^n \times {\det}^{-\frac{1}{2}}\l \frac{ V\mathbf{M}_n}{2\pi} \r,
\ee
with $\mathbf{M}_n$ the $2n\times 2n$ matrix
$\mathbf{M}_n$ 
\be\label{eq:Matrix_M}
\begin{split}
\mathbf{M}_n =  \l\frac{2}{\tilde{\tau}^2} +\varepsilon_2\r\mathds{1} + \varepsilon_2(1-V_A/V)
\begin{pmatrix}
  0 & \rvline & -\mathds{1} \\
  \hline
  -\mathds{1} & \rvline & 0
\end{pmatrix} + \varepsilon_2 V_A/V
\begin{pmatrix}
  \begin{matrix}
  0 & \dots & \dots & 0 \\
  \dots & \dots & \dots & \dots\\
  \dots & \dots & \dots & \dots\\
  0 & \dots & \dots & 0
  \end{matrix}
  & \rvline & 
  \begin{matrix}
  0 & -1 & 0 & \dots \\
  \dots & 0 & -1 & \dots\\
  \dots & \dots & \dots & \dots\\
  -1 & 0 & 0 & \dots
  \end{matrix}
  \\
\hline
  \begin{matrix}
  0 & \dots & \dots & -1 \\
  -1 & 0 & \dots & 0\\
  0 & -1 & \dots & 0\\
  \dots & \dots & \dots & \dots
  \end{matrix}
   & \rvline &
  \begin{matrix}
  0 & \dots & \dots & 0 \\
  \dots & \dots & \dots & \dots\\
  \dots & \dots & \dots & \dots\\
  0 & \dots & \dots & 0
  \end{matrix}
\end{pmatrix},
\end{split}
\ee
whose spectrum is thoroughly discussed in Sec.~\ref{app:Diag_mat}. Finally, we compute the difference of R\'enyi entropies between $\ket{\Psi_\tau}$ and $\ket{\Psi_0}$ as
\be\label{eq:Sn_sh_filter}
S_{n}(A)  = S_{n,0}(A) + \frac{1}{2(n-1)}\log \frac{\det(\mathbf{M}_n)}{\det^n(\mathbf{M}_1)}.
\ee
For small $\tilde{\tau}$ one gets a quadratic growth as $S_n(A) - S_{n,0}(A) \sim \tilde{\tau}^2$. For large $\tilde{\tau}$ instead a logarithmic growth is observed
\be\label{eq:log_entropy}
S_n(A) - S_{n,0}(A) \simeq \log \tilde{\tau} + \dots,
\ee
which will be traced back to the smallest eigenvalue of $\mathbf{M}_n$. As expected, our prediction vanishes identically in the limit $V_A/V \rightarrow 0$; this is a further demonstration that local observables, associated with a finite region $A$, do not vary in the short filter time regime (see Eq.~\eqref{eq:Sn_sh_filter}). Further, one can check explicitly that the result is symmetric under $V_A \leftrightarrow V-V_A$, as expected since $\ket{\Psi_\tau}$ is a pure state.

In the case of a medium filter time, with $\tau$ fixed, one can still perform a saddle-point analysis, but carefulness is needed. In particular, $\lambda_\tau(t_j)$ (and $\lambda_\tau(\tilde{t}_j)$) no longer contribute to the saddle point of \eqref{eq:Exp_twist}, and the latter is determined by the term in Eq.~\eqref{eq:Exp_twist1}. Here localization of the integral around $t_j \simeq \tilde{t}_j \simeq t$, which is a one-dimensional manifold, occurs. To compute \eqref{eq:Exp_twist}, we first perform a saddle-point integration over the transverse ($2n-1$) directions and, then, integrate over $t$
\be\label{eq:Exp_twist_med}
\begin{split}
^{n}\bra{\Psi_\tau}\mathcal{T}_A\ket{\Psi_\tau}^{n} =  \frac{\int dt |\lambda_\tau(t)|^{2n} \ {}^n\bra{\Psi_0}\mathcal{T}_A(t)\ket{\Psi_0}^n}{\l\int dt |\lambda_\tau(t)|^2\r^n} \times \frac{
\int d\mathbf{t} \exp \l -\frac{1}{2}V \mathbf{t}^T \mathbf{N}_n \mathbf{t}\r
}{\l\int dt' \exp \l -\frac{1}{2}V t' \mathbf{N}_1 t'\r\r^n} = \\
\frac{\int dt |\lambda_\tau(t)|^{2n} \ {}^n\bra{\Psi_0}\mathcal{T}_A(t)\ket{\Psi_0}^n}{\l\int dt |\lambda_\tau(t)|^2\r^n} \times \l \frac{V}{2\pi}\r^{\frac{1-n}{2}}\times \frac{\text{det}^{-\frac{1}{2}}\l \mathbf{N}_n\r}{\text{det}^{-\frac{n}{2}}\l \mathbf{N}_1\r}
\end{split}
\ee
where $\mathbf{t}$ is the $2n-1$-dimensional vector
\be
\mathbf{t} = \begin{pmatrix}t_1\\ \dots\\ t_{2n-1}\end{pmatrix},
\ee
and $\mathbf{N}_n$ is a $2n-1 \times 2n-1$ matrix. The specific entries of $\mathbf{N}_n$ depend on the way we parametrize the $2n-1$-dimensional manifold that is orthogonal to the one-dimensional one where the integral localizes. However, the spectrum of $\mathbf{N}_n$ does not depend on this choice and it can be obtained directly from the results available for $\mathbf{M}_n$ in Sec.~\ref{app:Diag_mat}. For instance, in the limit $\tilde{\tau}\rightarrow +\infty$ in Eq.~\eqref{eq:Matrix_M}, $\mathbf{M}_n$ becomes singular as one eigenvalue vanishes: this eigenvalue is associated precisely with the one-dimensional manifold of localization ($t_j=\tilde{t}_j = t$ in Eq.~\eqref{eq:Exp_twist}), while the other ($2n-1$) ones rule the exponential decay in its neighborhood. The latter are the eigenvalues of $\mathbf{N}_n$, and we write the spectrum of the matrix (denoted by $\text{Spec}(\dots)$) as
\be
\text{Spec}\l \mathbf{M}_n|_{\tilde{\tau}=\infty}\r = \text{Spec}\l \mathbf{N}_n\r \cup \{0\}.
\ee
For completeness, we write the eigenvalues of $\mathbf{N}_n$ as
\be\label{eq:eigs_Nn}
\begin{split}
 \{ \varepsilon_2( 
 1+\sqrt{(1-V_A/V)^2 + (V_A/V)^2 + 2 V_A/V(1-V_A/V)\cos k})\}_{k\geq 0}\cup\\
\{ \varepsilon_2( 
 1-\sqrt{(1-V_A/V)^2 + (V_A/V)^2 + 2 V_A/V(1-V_A/V)\cos k})\}_{k> 0},
 \end{split}
\ee
with $k\in \frac{2\pi}{n} \{0,\dots,n-1\}$.

To proceed with the analysis, it is necessary to make some hypothesis on the behaviour of ${}^n\bra{\Psi_0}\mathcal{T}_A(t)\ket{\Psi_0}^n$, which is the only quantity not predicted by this approach and depends on the properties of the model. We assume that the growth in time of the entropy under unitary dynamics from the state $\ket{\Psi_0}$ is linear, which is a common scenario found for both integrable and non-integrable systems \cite{cc-05,zn-20,clm-16}. At large times, the entropy is expected to approach an extensive value, and hence 
\be\label{eq:decay_twist}
\begin{split}
&\frac{{}^n\bra{\Psi_0}\mathcal{T}_A(t)\ket{\Psi_0}^n}{{}^n\bra{\Psi_0}\mathcal{T}_A(0)\ket{\Psi_0}^n} \simeq \begin{cases}
\exp\l -(n-1)\Gamma_n|\partial A|\times |t| \r \quad |t|\lesssim t_{\text{th}},\\
\exp\l -(n-1)s_n V_A\r \quad |t|\gtrsim t_{\text{th}},
\end{cases}
\end{split}
\ee
with $\Gamma_n\geq 0$ a (model-dependent) rate, and $|\partial A|$ the area of $A$. The thermalization time $t_{\text{th}}$ entering the previous expression grows with the size of the region and it is proportional to $t_{\text{th}}\sim V_A^{1/d}$ in the presence of ballistic transport (or $t_{\text{th}}\sim V_A^{2/d}$ for diffusive systems). For one-dimensional systems, where $\partial A$ is just a set of points (which means that $|\partial A|$ does not scale with the subsystem size) the first term in Eq.~\eqref{eq:Exp_twist_med} is finite, as long as $\tau$ is fixed and much smaller than the thermalization time $t_{\text{th}}$. For those systems, the leading term of the entropy is precisely given by the second term of Eq.~\eqref{eq:Exp_twist_med}
\be\label{eq:Sn_med}
S_n(A)-S_{n,0}(A) \simeq \frac{1}{2}\log V_A + \dots
\ee
where $O(1)$ ($\tau$-dependent) terms have been neglected\footnote{While \eqref{eq:Sn_med} is derived within the hypothesis of $V_A/V$ fixed, it also holds in the limit $V_A/V \rightarrow 0$. One can show it via a careful analysis of the spectrum of $\mathbf{N}_n$, reported in Eq.~\eqref{eq:eigs_Nn}, in the limit above}.

We now discuss the limit of large $\tau$ for the corrections to Eq.~\eqref{eq:Sn_med}. We analyze the cases $n\geq 2$, $n\rightarrow 1^+$, and $n<1$ separately, as qualitative differences arise. The important quantity is the first term of Eq.~\eqref{eq:Exp_twist_med}, and we study its behaviour as a function of $\tau$. First, using the exponential decay in Eq.~\eqref{eq:decay_twist}, we make the following approximation
\be\label{eq:Integral_n}
\int dt |\lambda_\tau(t)|^{2n} \ \frac{{}^n\bra{\Psi_0}\mathcal{T}_A(t)\ket{\Psi_0}^n}{{}^n\bra{\Psi_0}\mathcal{T}_A(0)\ket{\Psi_0}^n} \simeq \int dt  \exp\l -(n-1)\Gamma_n|\partial A|\times |t| \r  \propto \frac{1}{(n-1)\Gamma_n |\partial A|}.
\ee
Also, up to an irrelevant prefactor, we have
\be
\l\int dt |\lambda_\tau(t)|^2\r^n \propto \tau^n.
\ee
Putting everything together, we express the leading correction to Eq.~\eqref{eq:Sn_med} as
\be\label{eq:Renyi_nbig1}
S_n(A)-S_{n,0}(A) \simeq \frac{1}{2}\log V_A + \frac{n}{n-1}\log \tau + \dots, \qquad n\geq 2
\ee
for large $\tau$. As manifest from the equation above, the analytic continuation $n\rightarrow 1$ is pathological, and the technical mechanism is traced back to the non-commutativity of the limits $\tau \rightarrow \infty$ and $n\rightarrow 1$. The leading order in the limit of small $(n-1)$ reads
\begin{align}
&\int dt |\lambda_\tau(t)|^{2n} \ \frac{{}^n\bra{\Psi_0}\mathcal{T}_A(t)\ket{\Psi_0}^n}{{}^n\bra{\Psi_0}\mathcal{T}_A(0)\ket{\Psi_0}^n} \simeq \int dt |\lambda_\tau(t)|^{2n} \ \l 1-(n-1)\Gamma_1 |\partial A| |t|+\dots\r \propto \\
&\frac{\tau}{\sqrt{n}}\l 1+(1-n)\mathcal{C}\Gamma_1 |\partial A| \tau + \dots\r,
\end{align}
where $\mathcal{C}$ is a dimensionless constant given by
\be
\mathcal{C} = \frac{\int dx \ |x|e^{-2x^2}}{\int dx e^{-2x^2}} = \frac{1}{\sqrt{2\pi}}.
\ee
We compute the von Neumann entropy from the limit $n\rightarrow 1$ of the expression above and get
\be\label{eq:lin_growth_vN}
S_1(A)-S_{1,0}(A) \simeq \frac{1}{2}\log V_A + \mathcal{C}\Gamma_1 |\partial A|\tau +\log \tau + \dots.
\ee
We finally consider the case $n<1$. This regime is particularly tricky, since the expectation value of the twist operator in Eq.~\eqref{eq:decay_twist} grows exponentially in time, and it competes with the decay of $\lambda_\tau(t)$ in Eq.~\eqref{eq:Exp_twist_med}. We perform an estimation via saddle point analysis, which gives the most leading asymptotics at large $\tau$ (up to an irrelevant constant prefactor)
\be
\begin{split}
\int^{\infty}_{-\infty} dt |\lambda_\tau(t)|^{2n} \ \frac{{}^n\bra{\Psi_0}\mathcal{T}_A(t)\ket{\Psi_0}^n}{{}^n\bra{\Psi_0}\mathcal{T}_A(0)\ket{\Psi_0}^n} \simeq &2\int^{\infty}_0 dt  \exp\l -\frac{2nt^2}{\tau^2} +(1-n)\Gamma_n |\partial A| t\r \propto\tau\exp\l \frac{(1-n)^2 \Gamma^2_n |\partial A|^2}{8n}\tau^2\r\, .
\end{split}
\ee
Putting everything together, we find
\be\label{eq:Renyi_n_small}
S_n(A)-S_{n,0}(A) \simeq \frac{1}{2}\log V_A + \frac{(1-n) \Gamma^2_n |\partial A|^2}{8n}\tau^2 +\log \tau + \dots \quad n<1, 
\ee
where subleading terms have been neglected. Remarkably, a quadratic growth emerges at large $\tau$, resulting in a much faster growth of the R\'enyi entropy compared to the von Neumann entropy, as described in Eq.~\eqref{eq:lin_growth_vN}, and the logarithmic behavior at $n>1$ in Eq.~\eqref{eq:Renyi_nbig1}.

We remark that the predictions above refer to $\tau$ large with respect to microscopic scales but still smaller than the thermalization time $t_{\text{th}}$. For instance, in the limit of $\tau \gg t_{\text{th}}$, the integral in \eqref{eq:Exp_twist_med} is dominated by the asymptotic value of ${}^n\bra{\Psi_0}\mathcal{T}_A(t)\ket{\Psi_0}^n$ (see Eq.~\eqref{eq:decay_twist}); that is exponentially small in the subsystem size, and therefore the R\'enyi entropy of $\ket{\Psi_\tau}$ becomes extensive. This is precisely the regime of long-filter time of the main text, where the thermal properties are eventually recovered.

\section{Diagonalization of the matrix $\mathbf{M}_n$}\label{app:Diag_mat}

\begin{figure}[t]
\includegraphics[width=0.5\columnwidth]{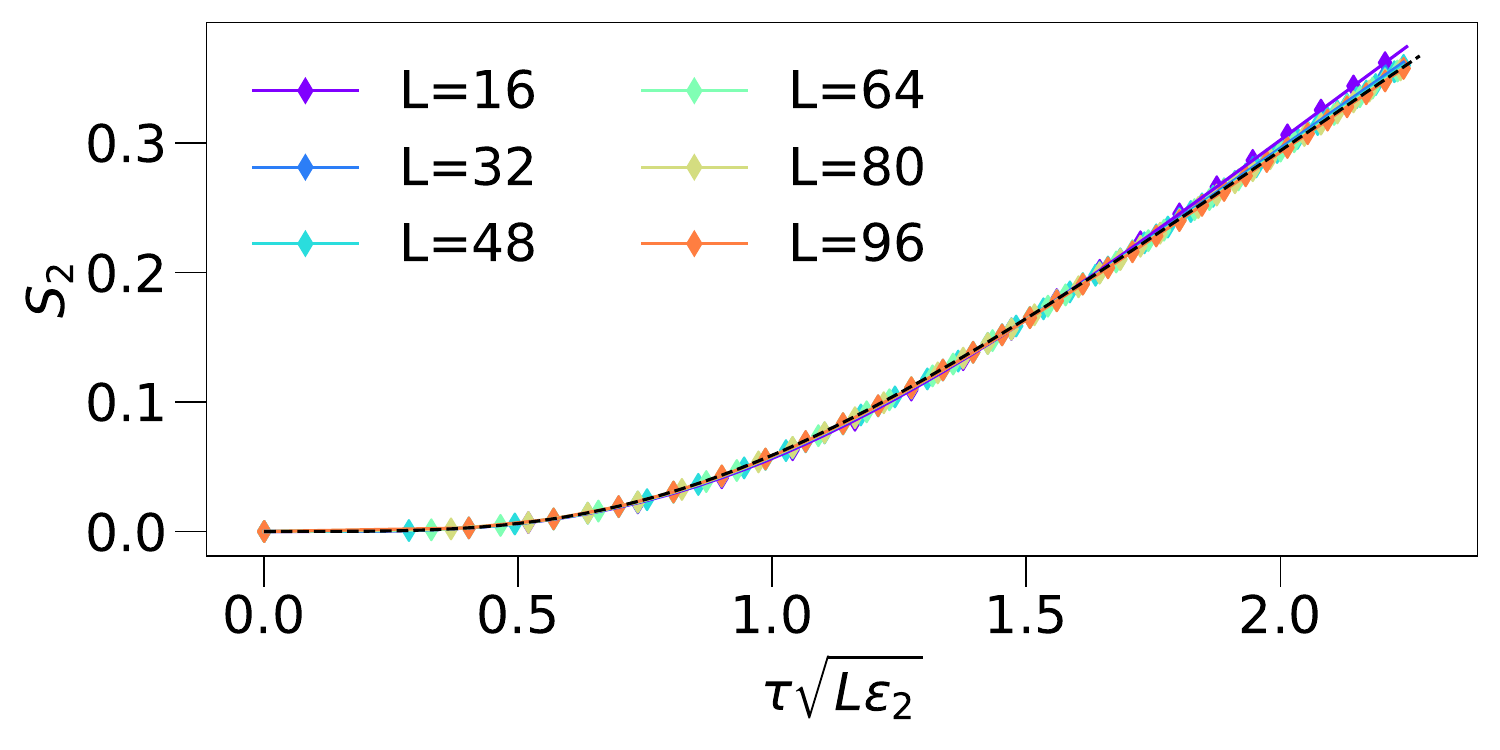}
\caption{R\'enyi entropy of half chain in the short filter time. The data shows a good collapse of $S_2$ against $\tau \sqrt{L\varepsilon_2}$ for various system sizes, and compatibility with the universal prediction \eqref{eq:S2_halfchain_formula}, black dashed line, is found.}\label{Fig:S2:SFT}
\end{figure}

Here, we diagonalize the matrix $\mathbf{M}_n$ defined in Eq.~\eqref{eq:Matrix_M} to provide close expressions for the entropies in the short filter time regime via \eqref{eq:Sn_sh_filter}. To do that, we make use of a $\mathbb{Z}_n$ symmetry associated with the replica permutation symmetry $j\rightarrow j+1$, which corresponds to $t_j\rightarrow t_{j+1},\tilde{t}_j \rightarrow \tilde{t}_{j+1}$ in Eq.~\eqref{eq:Exp_twist}. Such a symmetry allows us to decompose $M_n$ in $n$ $2\times$ blocks (via the Fourier transform) defined by
\be
\begin{split}
M(k) \equiv \l\frac{2}{\tilde{\tau}^2} +\varepsilon_2\r \begin{pmatrix} 1 & 0 \\ 0 & 1\end{pmatrix} +\varepsilon_2(1-V_A/V)\begin{pmatrix} 0 & -1 \\ -1 & 0\end{pmatrix} + \varepsilon_2 V_A/V \begin{pmatrix} 0 & -e^{ik} \\ -e^{-ik} & 0\end{pmatrix}.
\end{split}
\ee
Here, $k = 0, \frac{2\pi}{n}, \dots \frac{2\pi(n-1)}{n}$ correspond to the discrete momenta in the Fourier space. The diagonalization of  $M(k)$ is straightforward, and its two eigenvalues are
\be\label{eq:eigs_Mn}
\begin{split}
&\frac{2}{\tilde{\tau}^2} +\varepsilon_2\pm \varepsilon_2 \sqrt{(1-V_A/V)^2 + (V_A/V)^2 + 2 V_A/V(1-V_A/V)\cos k}.
\end{split}
\ee
Therefore, we express the determinant of $\mathbf{M}_n$ as
\be\label{eq:det_anal}
\text{det}\l \mathbf{M}_n\r  = \prod_k \text{det}\l \mathbf{M}(k)\r = \prod_k\l \l\frac{2}{\tilde{\tau}^2} +\varepsilon_2\r^2 -\varepsilon_2^2 \l (1-V_A/V)^2 + (V_A/V)^2 + 2 V_A/V(1-V_A/V)\cos k\r\r.
\ee
From the expression above, one can prove $\text{det}\l \mathbf{M}_n\r \geq \text{det}^n\l \mathbf{M}_1\r$, which ensures $S_n(A)\geq S_{n,0}(A)$ for $n\geq 2$ integer (Eq.~\eqref{eq:Sn_sh_filter}); this is physically expected since the filter protocol is supposed to increase the entropy of the state. For the sake of completeness, we exhibit the explicit analytic prediction of the R\'enyi-2 entropy of half of the system ($V_A/V=1/2$)
\be\label{eq:S2_halfchain_formula}
S_2(A)-S_{2,0}(A) \simeq -\log \left(\frac{2 \left(\varepsilon_2\tilde{\tau}^2+1\right)}{ \sqrt{\left(\varepsilon_2\tilde{\tau}^2+1\right) \left(\varepsilon_2\tilde{\tau}^2+2\right)^2}}\right),
\ee
valid in the large volume limit (with $\tilde{\tau} = \tau \sqrt{V}$ fixed). Fig. \ref{Fig:S2:SFT} shows a numerical check of Eq.~\eqref{eq:S2_halfchain_formula} for the non-integrable Ising model in Eq.~\eqref{eq:H_ising}.

We now discuss the analytic continuation of Eq.~\eqref{eq:det_anal} to non-integer values of $n$. We need the following trigonometric equality (see e.g. Appendix A3 of Ref.~\cite{cmc-22})
\be\label{eq:useful_identity}
\prod^{n-1}_{p=0}\l x- \cos \frac{2\pi p}{n} y\r = \l \l\frac{x+\sqrt{x^2-y^2}}{2}\r^{n/2} - \l\frac{x-\sqrt{x^2-y^2}}{2}\r^{n/2}\r^2,
\ee
and we express Eq. \eqref{eq:det_anal} as
\be\label{eq:det_anal1}
\text{det}\l \mathbf{M}_n\r = \l \l\frac{x+\sqrt{x^2-y^2}}{2}\r^{n/2} - \l\frac{x-\sqrt{x^2-y^2}}{2}\r^{n/2}\r^2,
\ee
with
\be
\begin{cases} x = \l\frac{2}{\tilde{\tau}^2} +\varepsilon_2\r^2 -\varepsilon_2^2 \l (1-V_A/V)^2 + (V_A/V)^2\r,\\
y= 2\varepsilon^2_2 V_A/V(1-V_A/V).
\end{cases}
\ee
From \eqref{eq:det_anal1}, one computes the R\'enyi entropies for non-integers values of $n$. We emphasize that the general logarithmic growth found in Eq. \eqref{eq:log_entropy} for integer $n\geq 2$ as a function of $\tilde{\tau}$ is also present for $n\rightarrow 1$ and $n<1$. This mechanism has to be contrasted with the drastic change of behaviour observed in the medium filter time as $n$ crosses the value $1$, discussed at the end of Sec.~\ref{app:Renyi_replica}.

\section{Numerical methods}\label{app:num_methods}

In this Section, we present (i) some further details on the numerical simulation that we discussed in the main text, and (ii) some additional numerical simulations that we performed in order to ensure the reliability of our results.
We recall that
our study focuses on 
the Ising model in Eq.~\eqref{eq:H_ising}; the initial state is the N\'eel state $\ket{\uparrow \downarrow \uparrow \downarrow \ldots \uparrow \downarrow}$, 
characterized by a vanishing energy density in the middle of the spectrum and an extensive energy variance.

Our numerical simulations are based on matrix-product states (MPS), a class of many-body quantum states that are characterized by a limited entanglement entropy and that display finite correlations decaying asymptotically exponentially in space in large enough one-dimensional systems~\cite{cpsv-21}.
The crucial parameter of MPS is the so-called bond link $\chi$: for $\chi=1$ MPS are product states, whereas for increasing $\chi$ they can accommodate for larger correlations and eventually, for large enough $\chi$, MPS can cover the entire Hilbert space of a finite-length quantum spin chain.
MPS are a crucial tool for the numerical simulation of one-dimensional quantum many-body systems since for most situations of interest it is possible to accurately describe the quantum state with a limited value of $\chi$ and thus at a tractable numerical complexity.

The numerical simulation of an energy-filtered quantum state is complicated by the fact that while $H$ is a local Hamiltonian, the operator $H^2$ is non-local, and standard techniques such as TEBD or tDMRG~\cite{schollwock2011_review, Vidal03, Vidal04, White04,Daley04} cannot be straightforwardly employed. 
For this reason, we use the Time-Dependent Variational Principle (TDVP) to implement the energy filter protocol~\cite{review_TDPV, Cirac_1TDVP, Haegeman_2TDVP} employing the ITensors library~\cite{ITensor, ITensor-r0.3}. 
In general, a TDVP is a scheme that projects the Schr\"odinger equation dictating the time-evolution of the state onto the manifold of MPS with fixed maximum bond link $\chi_{\rm max}$.
The projection scheme can be done in several ways, and in particular allowing for the modification of only one site of the MPS (1-TDVP, first introduced in Ref.~\cite{Cirac_1TDVP}) or of two neighboring sites (2-TDVP, first introduced in Ref.~\cite{Haegeman_2TDVP}).
In general, the 1-TDVP suffers from the problem that it is not possible to increase the bond-link of the initial state during the time evolution, and thus requires an initial state represented by a sufficiently large bond-link in order to be able to describe the time-evolved state.
The 2-TDVP algorithm does not suffer from this difficulty, but requires in turn a more important computational complexity.
Following Ref.~\cite{Goto_TDVP}, we first apply the 2-TDVP 
and we evolve the initial state in time until a user-defined bond dimension is reached. 
Subsequently, we employ the 1-TDVP algorithm.
This approach is a compromise between the computational efficiency, in terms of RAM and processing time, offered by 1-TDVP, and the mitigation of projection errors inherent in 2-TDVP, thereby enabling us to reach large filter time $\tau$ even for $L=80$ (see Ref. \cite{Goto_TDVP}). This is the technique employed to obtain the results presented in the main text, where
the maximal bond link is $\chi = 450$, the $\delta \tau = 0.01$ and $[J, h_x, h_z]=[1,1.2,0.8]$.

In order to probe the reliability of the TDVP algorithm, we have also performed simulations based on a straightforward expansion of the evolution operator, which is less efficient in terms of computational resources and time-step errors. 
This expansion is given by 
\be \label{eq:brutal_exp}
\exp\l -\frac{H^2 \tau^2}{4}\r = \left[\exp\left(-H^2\frac{\delta \tau^2}{4}\right)\right]^N \simeq \l\mathds{1}-H^2\frac{\delta \tau^2}{4}\r^N,
\ee
with $\tau = \sqrt{N}\delta \tau$ (valid in the large $N$ limit). 
Since the Hamiltonian $H$ has a simple MPO representation, the MPO representation of $H^2$ follows automatically, and thus the operator $\mathds{1} - \frac{\delta \tau^2}{4} H^2$ can be contracted onto an MPS with standard techniques; we set the cutoff to $10^{-10}$ for the contraction between the MPO and the MPS, and to $0$ for the construction of the MPOs.
Using this method we have been able to study spin chains up to $L=40$ and thus to produce numerical data to validate the TDVP data. We use $\delta \tau =0.1$ for $L=8,16,24$ and $\delta \tau =0.05$ for $L=32,40$.

\begin{figure}[t]
    \centering
    \includegraphics[scale=0.7, trim={0cm 0.5cm 0cm 0cm}, clip]{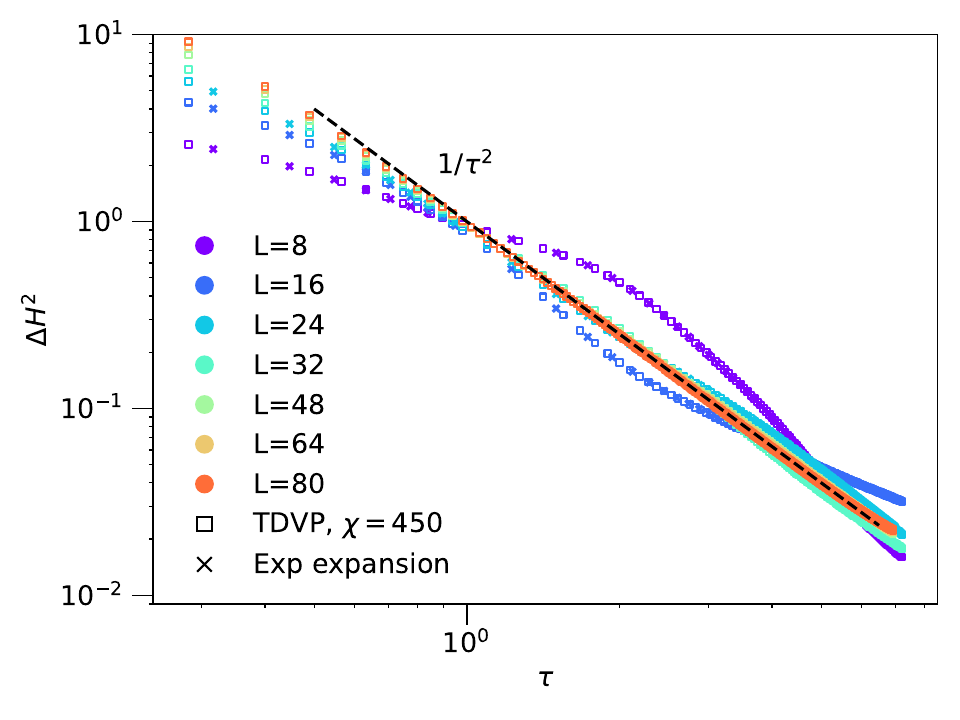}
    \caption{
    Energy variance $\Delta H^2$ as a function of $\tau$ for different system sizes. We have employed both the TDVP algorithm (bond dimension reached $\chi=450$, square markers) and the expansion \eqref{eq:brutal_exp} (crosses, for $L=8,16,24$), for $\delta \tau=0.1$. These methods are compatible, and they further show a collapse to $\Delta H^2 = \tau^{-2}$ at large $L$.   
   }
    \label{fig:var_MFT}
\end{figure}

In Fig.~\ref{fig:var_MFT} we plot several data for the energy variance of the model obtained with both techniques for $L$ up to $80$. The agreement of the two techniques is excellent up to $L=24$ and with small differences at $\tau > 5$ for $L=32, 40$; for larger system size we could not produce data with the expansion in Eq.~\eqref{eq:brutal_exp}.
For large $L$,
we compare our numerics with the analytical prediction, since in the large $L$ limit we know that as the filter time $\tau$ increases, it should decrease as $\Delta H^2 \simeq \tau^{-2}$. 
A collapse to this asymptotic scaling is observed for larger values of $L \sim 64, 80$, while clear finite size effects are present for $L=8, 16$. 
In this manner, we validate the results for both short and large $L$ obtained with the TDVP algorithm. 

\section{Further numerical studies} \label{app:Neel_Y+}

\begin{figure}
    \centering
    \includegraphics[scale=0.7, trim={0cm 0.5cm 0cm 0cm}, clip]{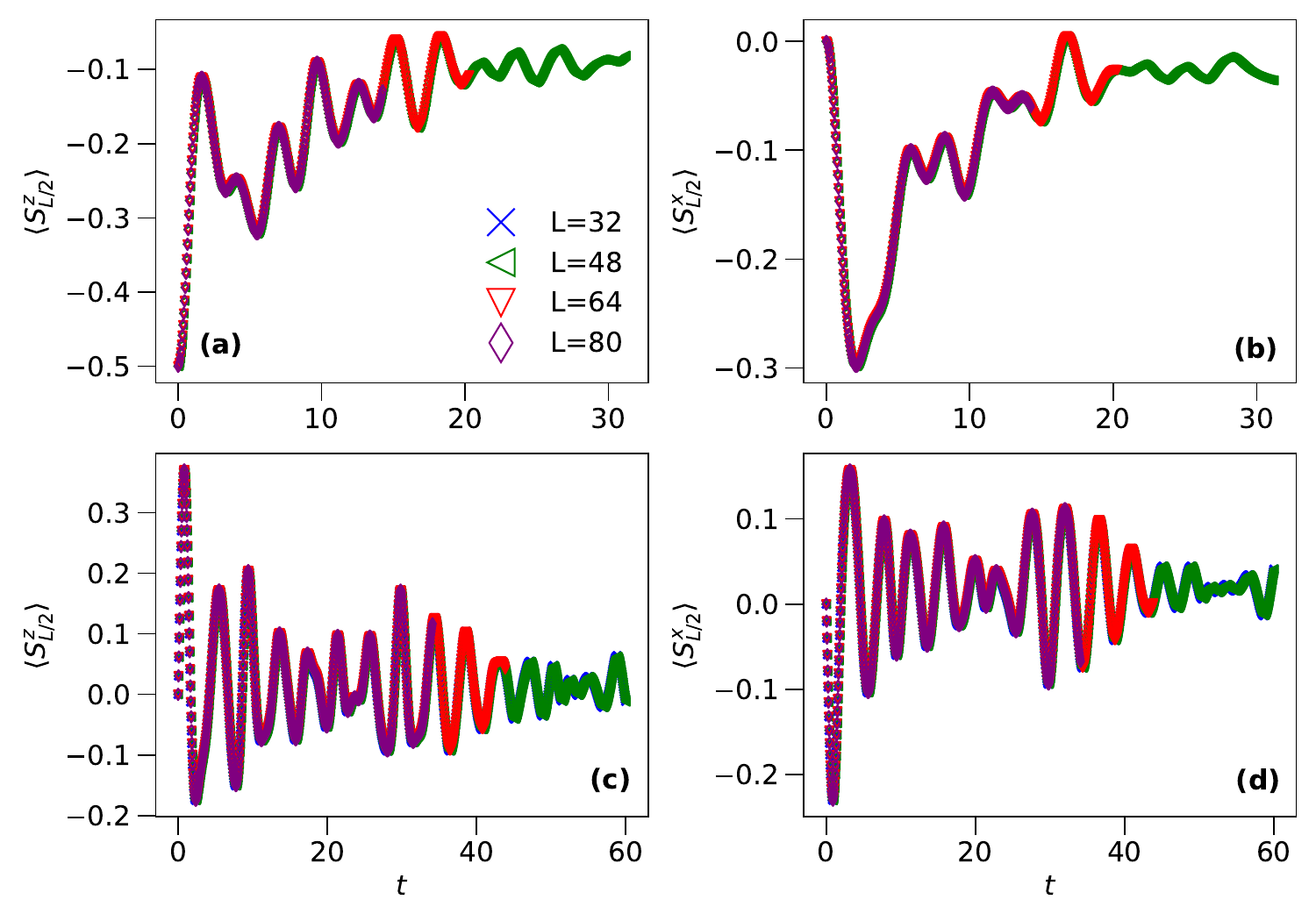}
    \caption{Unitary evolution of $\langle S^z_{j=\frac{L}{2}}\rangle$ and $\langle S^x_{j=\frac{L}{2}}\rangle$ as a function of $t$ for two different states: panels (a) and (b) show data for the N\'eel state, while panels (c) and (d) show data for the $\ket{\text{Y+}}$ state.}
    \label{fig:EV_unitary}
\end{figure}

The goal of this appendix is to provide numerical evidence of thermalization at large real times for the Ising model of the main text; in particular we focus on the initial N\'eel state and on the $\ket{\text{Y+}}$ state, in which all spins are aligned along the positive y direction.
The study of the unitary evolutions of the two states is in Fig.~\ref{fig:EV_unitary}; it is possible to see that the relaxation to the thermal value for the N\'eel state (panels (a) and (b)) is slow and it requires timescales larger than those considered in Fig.~1(c) of the main text. 
For the $\ket{\text{Y+}}$ state (panels (c) and (d)), the relaxation appears to be faster but with important oscillations, as also thoroughly studied in Ref.~\cite{Y+_Therm}.

\begin{figure}[t]
    \centering
    \includegraphics[scale=0.7, trim={0cm 0.5cm 0cm 0cm}, clip]{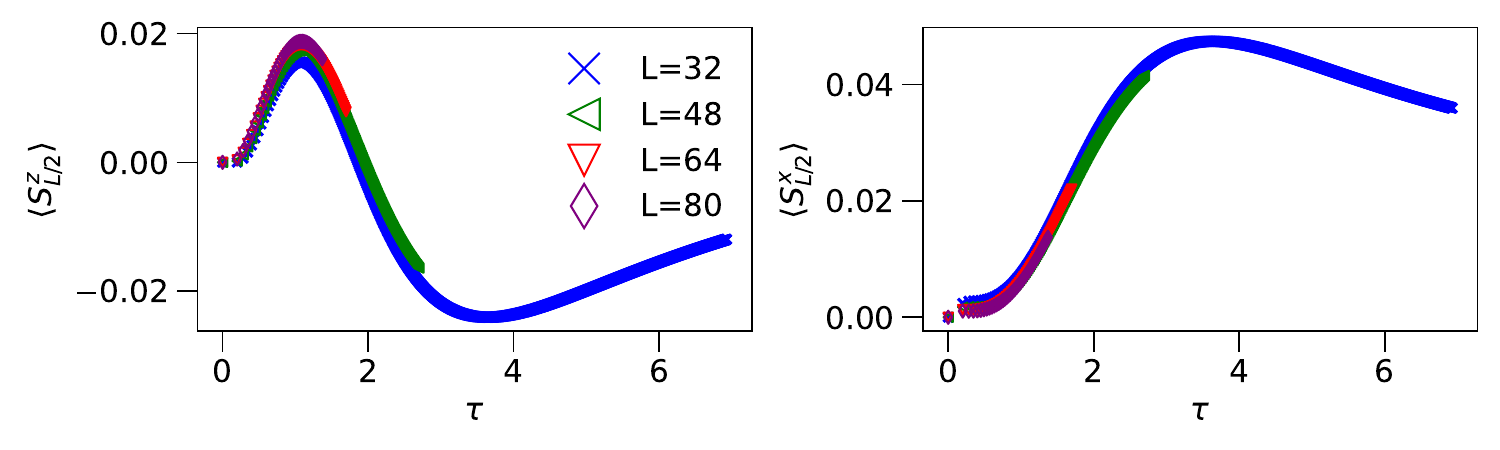}
    \caption{Filtered evolution for the $\ket{\text{Y+}}$ state of $\langle S^z_{j=\frac{L}{2}}\rangle$ and $\langle S^x_{j=\frac{L}{2}}\rangle$, as function of $\tau$. }
    \label{fig:Y+_filter}
\end{figure}

These data are interesting because they explain the fact that the filtering dynamics studied in the main text, up to $\tau \sim 7$, does not converge to the thermal expectation value.
In fact, also the filtering dynamics of the state $\ket{\text{Y+}}$, displayed in Fig.~\ref{fig:Y+_filter},  is clearly not at convergence at $\tau \sim 7$.

These results explain why thermal features are not observed for these two  states up to the medium filter time regime.
Since both states have an initial vanishing energy density, we expect that the observables that we analyzed ($\langle S^z_{j}\rangle$ and $\langle S^x_{j}\rangle$ at site $j=\frac{L}{2}$), assume, in the limit of $\tau \to\infty$, the value of the infinite-temperature state, that in our case is zero. This limit, due to the actual limitations of the numerical methods, is impossible to be verified; we are able only to reach $\tau \sim 7$ and in this regime, the filtered dynamics is not able to describe the thermal value. 

For completeness, we have used a bond dimension of $250$ and $0.05$ as time-step to simulate the unitary evolutions; while for the filtered evolution the bond dimension and the other parameters are the same employed in Appendix~\ref{app:num_methods}.

\end{document}